\documentclass[preprint]{aastex63}
\usepackage{color}

\accepted{June 28, 2020}
\shorttitle{Magnetic Reconnection in Relativistic Current Sheet}
\shortauthors{Hoshino}
\begin{document}
\title{Stabilization of Magnetic Reconnection in Relativistic Current Sheet}
\correspondingauthor{Masahiro Hoshino}
\email{hoshino@eps.s.u-tokyo.ac.jp}
\author[0000-0002-1818-9927]{Masahiro Hoshino}
\affil{Department of Earth and Planetary Science,
The University of Tokyo, Tokyo 113-0033, Japan}
\begin{abstract}
Recently, magnetic reconnection, where a relativistically hot plasma is confined by a strong magnetic field, has received great attention in relation to astrophysical objects (e.g., pulsar magnetosphere and magnetar). However, reconnection with relativistic high-speed drift current in the plasma sheet has not been investigated yet. Thus, from both the theoretical and computational points of view, we studied the growth rate of relativistic reconnection for the high-speed drift current in a relativistically hot plasma sheet. Consequently, we argue that, contrary to the conventional understanding of the fast energy dissipation by the relativistic reconnection, the growth of magnetic reconnection is suppressed.
\end{abstract}

\section{Introduction}
Rapid electromagnetic energy dissipation and particle acceleration in the relativistic magnetic reconnection have been getting much attention because many astrophysical objects are known to store high quantities of magnetic energy in their system and to rapidly release such stored magnetic energy. The pulsar magnetosphere, magnetar, and active galactic nucleus (AGN) jet are manifestations of such rapid release of huge electromagnetic energy. Over the last few decades, the dynamics and particle acceleration of the relativistic reconnection have been extensively investigated by many researchers \citep[e.g.][]{Zenitani01,Hoshino12,Guo14,Sironi14,Uzdensky16,Blandford17}.
It has been argued that magnetic reconnection with a relativistic plasma temperature $T/mc^2 > 1$ can quickly release the stored magnetic energy and generate non-thermal particles with a hard power-law energy spectrum of $N(\varepsilon) \propto \varepsilon^{-p}$ with $p>1$, where $mc^2$ is the rest mass energy of particle and $\varepsilon$ is the particle energy \citep{Zenitani01,Jaroschek04}.  

The relativistic reconnection used to be parameterized by the magnetization parameter $\sigma = B^2/4 \pi N m c^2$, the ratio of the available magnetic field energy to the particle rest mass energy. With regard to an anti-parallel magnetic field configuration (e.g., Harris equilibrium), the magnetic pressure $B^2/8 \pi$ outside the plasma sheet balances with the plasma gas pressure $P=NT$ inside the plasma sheet, and $\sigma$ is then regarded as a parameter for the characterization of the plasma temperature confined in the plasma sheet. Since the Alfv\'{e}n velocity is simply given by $V_A/c = \sqrt{\sigma/(\sigma+1)}$ in the cold plasma limit, the reconnection jet may be accelerated up to almost Alfv\'{e}n velocity by the tension force of the reconnecting magnetic field line. Therefore, generally speaking, the reconnection outflow speed in the exhaust region may reach up to almost the speed of light for $\sigma \gg 1$. Moreover, in relation to the high-speed reconnection jets, the strong electric field $E$, whose magnitude becomes almost equal to the reconnecting magnetic field $B$, can be induced. The stronger generated $E$ leads to a more efficient particle acceleration. Recently, the $\sigma$ dependence on particle acceleration during reconnection has been extensively investigated by means of particle-in-cell (PIC) simulations \citep[e.g.][]{Sironi14,Guo14}. \citet{Sironi14} reported that a hard energy spectrum whose spectral index is smaller than $2$ can be obtained for plasma sheet with a large $\sigma >10$.

So far, it has been understood that the relativistic reconnection with $\sigma >1$ is a kind of efficient engine for releasing stored magnetic field energy into the kinetic energy of particle. However, the drift velocity in the current sheet is another important parameter prescribing the relativistic reconnection. Nevertheless, the effect of the drift velocity has not yet been seriously considered. The drift velocity $u_d$ under the charge neutral plasma can be determined by the force balance between the pressure gradient force $-\nabla P$ and the Lorentz force $eN u_d B/c$, while the drift speed can be obtained by $u_d/c \sim T/\lambda e B \sim r_L/\lambda$, where $\lambda$ and $r_L$ are the thickness of the plasma sheet and the relativistic gyro-radius of particle, respectively. Therefore, when the thickness of the plasma sheet becomes as thin as the gyro-radius, the drift velocity $u_d$ becomes relatively close to the speed of light $c$.

In fact, the drift velocity may be close to the speed of light in some astrophysical settings. A pulsar wind, for example, may have a large drift velocity in the so-called strip current sheet \citep{Kirk03}. For a pulsar magnetosphere whose magnetic fields are extended from the surface of the neutron start, the toroidal magnetic field component may dominate as the pulsar wind/plasma is propagating outward. If the central neutron star is characterized by an oblique rotator with a finite angle between the axis of the dipole magnetic moment and the pulsar rotation axis, the toroidal magnetic field component in the equatorial plane changes its polarity during every rotation period. Thereafter, the current sheet is expected to be formed in the pulsar wind. While the pulsar wind is known to expand radially with a relativistic wind speed, the toroidal magnetic field $B$ decreases with the distance from the central pulsar (i.e., $B \propto 1/r$) and the plasma density is $N \propto 1/r^2$. Therefore, the drift velocity, which carries the electric current, increases with the distance from the pulsar.
As such, the drift velocity $u_d$ in the proper frame may be given by,
\begin{equation}
  \frac{u_d}{c} = \frac{(B/\Gamma_{\rm wind})}{8 \pi (\delta \Gamma_{\rm wind}) e (n/\Gamma_{\rm wind})}
  = \left( \frac{B_{lc}}{8 \pi \delta e N_{\rm GJ} \kappa \Gamma_{\rm wind}} \right)
  \left( \frac{r}{r_{lc}} \right),
\label{eq:J_drift}
\end{equation}
where $r_{lc}$ is the radius of the light cylinder $c/\Omega$, and $B_{lc}$, $\delta$, and $\Gamma_{\rm wind}$ are the magnetic field at $r=r_{lc}$, the thickness of the current sheet, and the bulk Lorentz factor of the expanding wind in the observer frame, respectively.  Note that $B$ and $n$ ($\delta$) in the observer frame become $\Gamma_{\rm wind}$ times larger (smaller) than those in the proper frame.
$N_{\rm GJ}=\Omega \cdot B / 2 \pi e c$ is the Goldreich-Julian density \citep{Goldreich69} and $\kappa \sim 10^3-10^4$ is the multiplicity of the charged particles \citep{Arons83}.
We have assumed $u_d = u_{di}=-u_{de}$, where $u_{di}$ and $u_{de}$ are the drift velocities of the positive and negative charged particles, respectively.
The thickness of the current sheet $\delta$ is smaller than $c/\Omega$, and we set $\delta=\varepsilon c/\Omega$ with $\varepsilon < 1$. Thereafter, we obtain
\begin{equation}
  \frac{u_d}{c} = \frac{1}{4 \kappa \varepsilon \Gamma_{\rm wind}} \frac{r}{r_{lc}}.
\label{eq:drift_vel}
\end{equation}
As the distances of the light cylinder $r_{lc}$ and of the termination shock $r_{sh}$ are believed to be $r_{lc} = c/\Omega \simeq 10^9 \rm{cm}$ and $r_{sh}=0.1~\rm{pc} \sim 3 \times 10^{18} \rm{cm}$ for the crab pulsar wind, respectively, we can easily find that $u_d$ in the pulsar wind may reach up to the speed of light before attaining the position of the termination shock when $\varepsilon \kappa \Gamma_{\rm wind} < 10^9$.  
The regime of reconnection that has a high-speed drift current seems to appear in the pulsar wind.

So far, it seems to be simply believed that a higher drift velocity set up in the plasma sheet leads to a faster reconnection because the thin current sheet with the higher drift velocity has larger free energy to destabilize the current sheet. Therefore, no investigation has been done for the reconnection with a high-speed drift current.
However, the conductivity of collisionless reconnection, which plays an important role on magnetic energy dissipation, may be modified in the relativistic drift current when the drift velocity becomes close to the speed of light. The collisionless conductivity is provided by the resonance process between the (inductive) reconnection electric field and the particles accelerated by its electric field in and around the X-point. In line with this, the special relativity effect during the resonance process should be considered. Accordingly, the growth rate of reconnection may be modified. In this study, we investigate the effect of the high-speed drift current on the relativistic magnetic reconnection and argue that the growth of the reconnection can be significantly reduced as the drift current speed approaches to the speed of light.

\section{Comparison between PIC simulation and the standard theory of relativistic tearing instability}

Before investigating the relativistic reconnection with a high-speed drift current sheet, let us quickly review the standard theory of the relativistic tearing mode discussed by \cite{Zelenyi79}, where the plasma temperature is relativistically hot, but the drift speed $u_d = \beta c$ is assumed to be non-relativistic, i.e., $\beta \ll 1$. In this framework, the growth rate is given by
\begin{equation}
\gamma \tau_c \simeq \frac{2 \sqrt{2}}{\pi} k \lambda (1-k^2 \lambda^2) \beta^{3/2},
\label{eq:growthrateLev}
\end{equation}
where $\tau_c = \lambda/c$ and $k$ are the light transit time across the plasma sheet and the wave number, respectively.

In this theory, we can see that the growth rate $\gamma$ increases with increasing the drift speed $\beta$. The same characteristic of the dependency of the growth rate on the drift velocity is also seen in the non-relativistic tearing instability whose plasma temperature is much less than the rest mass energy \citep{Coppi66,Hoh66}. The free energy of the localized electric current in the current sheet rises with the increasing $\beta$. Thereafter, we simply assumed that the reconnection rate increases with the drift speed.

However, if the drift speed becomes close to the speed of light, the relativistic plasma effect should be considered. In a collisionless tearing mode, the magnetic energy dissipation is provided by the Landau resonance between the inductive electric field associated with the change of the magnetic field and the particles energized by the inductive electric field. This magnetic energy dissipation rate through Landau resonance may be modified in the regime of the relativistic drift speed due to the effects of the special relativity, such as the retarded time and the relativistic inertia of particle.

Let us first study the time evolution of the tearing mode instability using a 2D PIC simulation with a periodic boundary in the $x$ direction and the conducting walls for the upper and lower boundaries at $y= \pm 16 \lambda$, where $\lambda$ is the thickness of the initial plasma sheet. The total system size is $64 \lambda \times 32 \lambda$ and the computational grid size is $1600 \times 800$. The total number of particles is $4.5 \times 10^8$ to calculate the high $\beta$ plasma sheet as accurately as possible.

As an initial equilibrium state, for simplicity, we adopted the Harris solution \citep{Harris62} for the pair plasma with the same mass and temperature for positron and electron, i.e., $m=m_{+}=m_{-}$ and $T = T_{+}=T_{-}$, respectively.
The magnetic field $B$ and the plasma density $N$ in the laboratory frame are given by
\begin{equation}
  B_x(y) = B_0 \tanh( \frac{y}{\lambda}),
\end{equation}
and
\begin{equation}
  N(y)=N_0 \cosh^{-2}( \frac{y}{\lambda}) + N_b,
\end{equation}
where $N_b$ is the background uniform plasma density. In addition, we have the relationship of the pressure balance as $B_0^2/ 8 \pi = 2 N_0 T$ and the force balance as $2 T/\lambda = |e| \beta B_0$.

The corresponding velocity distribution functions $f_{\pm}({\bf p})$ for positron $(+)$ and electron $(-)$ are set up by
\begin{equation}
  f_{\pm}({\bf p}) = \frac{\bar{N}_0}{4 \pi m^2 c \Theta K_2(mc^2/\Theta)}
  \exp \left(-\frac{E \mp c \beta p_z}{T}   \right),
\label{eq:Harris_f}
\end{equation}
where $E=\sqrt{m^2 c^4 +p^2c^2}$ and $T=\Theta/\Gamma_{\beta}$ are the particle energy and the plasma temperature in the laboratory frame, respectively. $\Gamma_{\beta}=1/\sqrt{1-\beta^2}$ is the Lorentz factor of the drift velocity. $\bar{N}_0 = N_0/\Gamma_{\beta}$ is the plasma density in the plasma rest frame. $K_2$ is a modified Bessel function \citep[e.g.][]{Kirk03}.

In our simulation study, the plasma temperature $T/mc^2=10$, the magnetization parameter $\sigma=20$, and the thickness of plasma sheet $\lambda=25$ grid size have been kept. The background plasma density $N_b$ is set to be $5 \%$ of the maximum Harris plasma density $N_0$ in the plasma sheet. We studied the growth rate of the tearing mode in association with the formation of magnetic islands as the function of the drift velocity $\beta c$.

\begin{figure}
\plotone{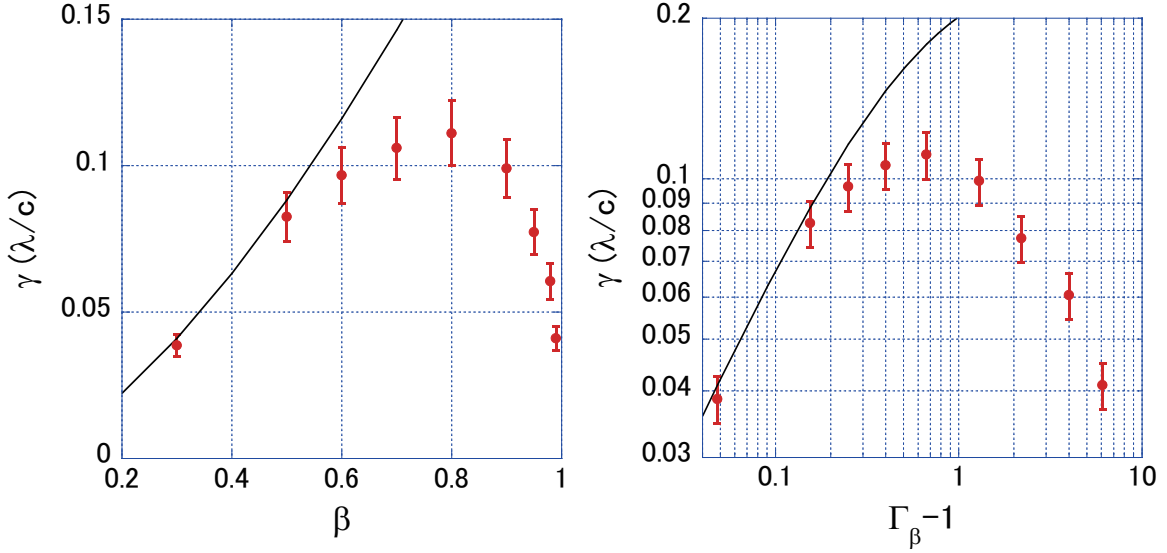}
\caption{Growth rates of relativistic tearing instability as function of the drift velocity $\beta$.  The left-hand panel is depicted in double-linear scales, and the right-hand panel is in double-logarithmic scales and the horizontal axis is $\Gamma_{\beta}-1$.  The red circles are the simulation results, and the solid lines are the theoretical curve given by \citet{Zelenyi79}, in which the relativistic drift effect is not taken into account.}
\label{fig:FIG1}
\end{figure}

The initial look of the time evolution of magnetic reconnection is basically same as those discussed in many other PIC simulations. We can see the formation of several magnetic islands with X- and O-points in the current sheet and the vortex motion of plasma around the islands (not shown here), suggesting that the tearing mode occurs in the relativistic current sheet. 
To measure the linear growth rates of the tearing mode from the time evolution, first, we took the average of the reconnecting magnetic field $B_y$ over $\pm 4$ grids in the $y$ axis across the neutral sheet. Second, we obtained the Fourier spectrum for $\langle B_y \rangle$ along the $x$ axis. By fitting the growth curve of each Fourier mode in time with the function of $\exp(\gamma t)$, we were able to measure the linear growth rates $\gamma$ as the function of the Fourier mode, i.e., the wave number $k$.

The unstable modes in the tearing instability are known to exist in $k \lambda \le 1$. In our simulation study, the smallest wavenumber $k_{\rm min}$ is set to be $k_{\rm min} \lambda = 25 \pi/800 \simeq 0.098$. Given that the unstable modes of the tearing instability appear in $k \lambda \leq 1$, we have almost $10$ unstable modes in the Fourier space. We found that the maximum growth rate appears around $k \lambda \sim 0.3-0.5$ and its corresponding wave mode number of $3 - 5$.

Fig. 1 presents the comparison between the growth rates obtained in the PIC simulations and those from the standard theory of Equation (\ref{eq:growthrateLev}). The red circles with the 10 \% error bars are the growth rates obtained after taking the average of three or four of the largest growth rates around $k \lambda \sim 0.39$ in the abovementioned procedure. Note that we multiplied an arbitrary constant with the order of unity by the theoretical growth rate of Equation (\ref{eq:growthrateLev}), and that the growth rate at $\beta=0.3$ is consistent with that of the previous simulation result \citep{Zenitani07}. The horizontal axes are $\beta$ (left) and $\Gamma_{\beta}-1 = 1/\sqrt{\beta^2 -1} -1$ (right).

As long as the drift speed $\beta$ is less than $0.5$, the linear theory given is consistent with our simulation results. However, we can see a large discrepancy between the linear theory and the simulation with regard to the larger $\beta$. The growth rate obtained by our simulations decreases with the increasing $\beta$ for $\Gamma_{\beta} > 1$.

We likewise studied the effect of plasma temperature on the linear growth rate and found that the growth rate is not dependent on the temperature for the relativistically hot plasma regime of $T/mc^2 > 1$. This result is consistent with Equation (\ref{eq:growthrateLev}) and has been already confirmed in the PIC simulation done by \citet{Zenitani07}.

\section{New linear theory of relativistic tearing instability}

Considering the foregoing comparison, let us extend the linear theory of the relativistic tearing mode to the regime of the relativistic high-speed drift current sheet with $\Gamma_{\beta} = 1/\sqrt{\beta^2 - 1} > 1$.
According to the standard procedure of the tearing mode theory \citep[e.g.][]{Coppi66,Hoh66,Zelenyi79},
the linearized Vlasov equation for a 2D tearing mode under the Harris distribution function of Equation (\ref{eq:Harris_f}) in the $x-y$ plane can be given by
\begin{equation}
\left( \frac{\partial}{\partial t} + 
  {\bf v} \cdot \frac{\partial}{\partial {\bf x}} + 
  \frac{e_j}{c}({\bf v} \times {\bf B}_0) \cdot \frac{\partial}{\partial {\bf p}} \right) f_{1j} 
=\frac{e_j f_{0j}}{c T_j} 
  \left(  (\frac{\partial}{\partial t} + {\bf v} \cdot \nabla) u_j A_{1z}
  -\frac{\partial}{\partial t}(v_z A_{1z}) \right),
\end{equation}
where $f_{j}({\bf x},{\bf v},t)$ is the velocity distribution function for a species $j$, the magnetic field ${\bf B} = \nabla \times {\bf A}$, and ${\bf p} = m_j {\bf v}/\sqrt{1-v^2/c^2}$. The suffixes of $0$ and $1$ represent the zero-th and the first order quantities, respectively.  The perturbation is represented by the frequency $\omega$ and the wave number $k$, and $A_{1z}(x,y,t)=\tilde{A}_{1z}(y) e^{ikx-i \omega t}$.
Note that the drift velocity of the electric current is parallel to the $z$ direction. We have assumed that $A_{1x}$ and $A_{1y}$ are small compared to $ A_{1z}$. After the time integration of the abovementioned equation along unperturbed particle trajectories from $-\infty$ to the current time $t$, we obtained
\begin{eqnarray}
f_{1j} &=& \frac{e_j f_{0j}}{c T_j} 
\left(  u_j A_{1z}+ i\omega \int_{-\infty}^{t} (v_z A_{1z}) dt'
\right), \nonumber \\
&=& \frac{e_j f_{0j}}{c T_j} 
\left(  u_j A_{1z}+ i\omega \int_{-\infty}^{t} 
v_z \tilde{A}_{1z}(y') e^{ikx'-i\omega t'} dt'
\right). 
\label{eq:tearingF}
\end{eqnarray}

The first term without the symbol of the time integration in Equation (\ref{eq:tearingF}) is referred to as the adiabatic/ideal MHD term, which represents the contribution from the magnetized motion of particles resulting from the slow change of the vector potential $A_1$. The second term involving the symbol of the time integration along unperturbed particle trajectories is identified as the non-adiabatic term, which represents the contribution from the departure from the foregoing adiabatic motion \citep[e.g.][]{Zelenyi79,Hoshino87}.  The main contribution of the second term comes from the meandering particle orbit near the neutral sheet, whose thickness $l_j$ is estimated as
\begin{equation}
l_j = \sqrt{r_{Lj} \lambda},
\end{equation}
where the gyro-radius is $r_{Lj} \sim T_j/eB$ for a relativistic hot plasma \citep[e.g.][]{Sonnerup71,Zelenyi79}. 
As in the conventional assumption of the particle trajectory motion, we may assume that the meandering motion can be treated as a type of free motion with a straight orbit,
i.e., $x'-x=v_x(t'-t)$ and $y'=y$. 
Thereafter, the time integration of the above equation becomes
\begin{eqnarray}
f_{1j} &\simeq& \frac{e_j f_{0j}}{c T_j} 
\left(  u_j A_{1z}+ i\omega \int_{-\infty}^{t} 
v_z \tilde{A}_{1z}(y) e^{ikx-i\omega t} e^{i(kv_x-\omega)(t'-t)} dt'
\right),  \nonumber \\
&\simeq& \frac{e_j f_{0j}}{c T_j} 
\left(  u_j A_{1z}+ i\omega 
v_z \tilde{A}_{1z}(y) e^{ikx-i\omega t} \int_{-\infty}^{0} e^{i(kv_x-\omega) \tau} d \tau
\right),  \nonumber \\
& \simeq & \frac{e_j f_{0j}}{c T_j} 
\left(  u_j A_{1z}- \frac{\omega}{\omega - kv_x + i0} v_z A_{1z} H(l_j)
\right),
\label{eq:tearingF2}
\end{eqnarray}
where $H(l_j)$ is a Heaviside function with $H(l_j)=1$ for $|y| \le l_j$ or $H(l_j)=0$.
The singularity of the denominator in the second term shows the Landau resonance, and represents the energy dissipation through the interaction between the electric field $E_{1z}=-(1/c)(\partial A_{1z}/\partial t)$ and the meandering particle.  As the tearing instability is the zero frequency mode with $\rm{Re}(\omega)=0$, the meandering particle with a small velocity $v_x$ can strongly resonate with the electric field.

As such, let us estimate the non-adiabatic electric current/resonance electric current.
By substituting the Harris equilibrium distribution function, we obtain
\begin{eqnarray}
J^{\rm res}_{1z}
&=& -\sum_{\pm} \int e_j v_z
\frac{e_j f_{0j}}{c T_j} \frac{\omega v_z}{\omega - kv_x + i0} A_{1z} d^3p, \nonumber \\
&=& -\sum_{\pm} \int \frac{e_j^2 \omega A_{1z}}{c T_j} 
\frac{\bar{N}}{4 \pi m^2 c \Theta K_2(mc^2/\Theta)} \exp \left(-\frac{\Gamma_{\beta}(E \pm c \beta p_z)}{\Theta}   \right)
\frac{v_z^2}{\omega - kv_x + i0} d^3p, \nonumber \\
&=& -\sum_{\pm} \int \frac{e^2 \omega A_{1z}}{c T_j} 
\frac{\bar{N}}{4 \pi m^2 c \Theta K_2(mc^2/\Theta)}
\Gamma_{\beta} (\bar{E} \pm \bar{p}_z \beta c) \exp \left(-\frac{\bar{E}}{\Theta} \right)
\frac{v_z^2}{\omega - kv_x + i0} \frac{d^3\bar{p}}{\bar{E}}, \nonumber
\end{eqnarray}
wherein we used the relationship of $d^3p/E = d^3\bar{p}/\bar{E}$.
For a relativistically hot plasma and/or a relativistic drift current sheet,
$v_z^2$ of the numerator may be approximated by $c^2$. Thus, we will get
\begin{eqnarray}
J^{\rm res}_{1z} 
&\simeq& -\sum_{\pm} \int \frac{e^2 \omega A_{1z}}{c T_j} 
\frac{\bar{N} \Gamma_{\beta}}{4 \pi m^2 c \Theta K_2(mc^2/\Theta)} \exp \left(-\frac{\bar{E}}{\Theta} \right)
\frac{c^2}{\omega - kv_x + i0} d^3\bar{p}, \nonumber \\
&\simeq& \sum_{\pm} \frac{e^2 \omega A_{1z}}{c T_j} 
\frac{\bar{N} \Gamma_{\beta}^2}{4 \pi m^2 c \Theta K_2(mc^2/\Theta)} \times 2 i \frac{\pi^2}{k} m^3 c^4
\exp \left( \frac{-mc^2}{\Theta} \right) \frac{\Theta}{mc^2}
\left(1 + 2 \frac{\Theta}{mc^2} (1+\frac{\Theta}{mc^2}) \right).
\label{eq:j_res0}
\end{eqnarray}
The main contribution of the resonance current comes from the Landau resonance with $\omega=k v_x$.  We have used the Plemelj relation to evaluate the Landau resonance term \citep{Dennery67}.
Considering $K_2(mc^2/\Theta) \sim 2 (\Theta/mc^2)$ for a relativistic hot plasma, the non-adiabatic/resonance current can be approximated by
\begin{equation}
J^{\rm res}_{1z}
\simeq \sum_{\pm} \frac{e^2 \omega A_{1z}}{c T_j} 
\frac{\bar{N} \Gamma_{\beta}^2}{4 \pi m^2 c} \times 2 i \frac{\pi^2}{k} m^2 c^2
\simeq \sum_{\pm} \frac{\pi}{2} \left( \frac{mc^2}{\Theta} \right)
\frac{\bar{N} e^2}{m} \frac{\Gamma_{\beta}^3}{kc} E_{1z}.
\label{eq:j_res}
\end{equation}

Inside the meandering layer, we found that the magnetic field is weak and the convection electric field $v \times B$ is negligible. After which, we were able to compare the above equation to the standard Ohm's law to estimate the effective conductivity for the relativistic tearing instability in collisionless plasma. The effective conductivity $\sigma_{eff}^{rel}$ in the relativistic regime can be expressed as
\begin{equation}
 \sigma_{\rm eff}^{\rm rel} = \frac{J^{\rm res}_{1z}}{E_{1z}} \simeq
 \frac{\bar{N} e^2}{m} \frac{\Gamma_{\beta}^3}{kc} \left( \frac{mc^2}{\Theta} \right)
 = \frac{N e^2}{m} \frac{\Gamma_{\beta}}{kc} \left( \frac{mc^2}{T} \right),
\label{eq:sigma_rel}
\end{equation}
where $N$ and $T$ without the bar symbol on the top of the character are the plasma density and the temperature in the laboratory frame, respectively.

We are now able to estimate the linear growth rate of the relativistic tearing instability with the standard procedure based on the energy principle. The change in magnetic field energy after integrating both $x$ and $y$ space with the periodic system in $x$ can be obtained by
\begin{equation}
\frac{\partial}{\partial t} \int_{- \infty}^{+ \infty} \frac{\bf B_1^2}{8 \pi} dy
= - \int_{- \infty}^{+ \infty} {\bf J_1} \cdot {\bf E_1^{*}} dy,
\label{eq:energy-pri-0}
\end{equation}
where ${\bf E_1^{*}}$ is the complex conjugate of ${\bf E_1}$ \citep[e.g.][]{Coppi66}.
In the energy principle, the term of the Poynting flux as $\int \nabla \times ({\bf E_1^{*}} \times {\bf B_1}) dy$ disappears due to $E_1=0$ at $y= \pm \infty$.

The perturbed electric current $J_1$ consists of two different parts, depending on the response to the electric field: one is the non-adiabatic current $J_{1}^{\rm res}$ given in the foregoing equation and the other is the adiabatic current $J_1^{\rm ad}$ that comes from the first term without Landau resonance in Equation (\ref{eq:tearingF2}). The solution of the adiabatic current is conventionally expressed as \citep{White77}
\begin{equation}
{\bf J_1^{\rm ad}} = \frac{c}{4 \pi} \nabla \times
(\nabla \times A_{1z}^{\rm ad}{\bf e_z}) =\frac{c}{4 \pi}
\frac{2A_{1z}^{\rm ad}(x,y,t)}{\lambda^2 {\rm cosh}^2(y/\lambda)} {\bf e_z},
\label{eq:ex-solution-tearing}
\end{equation}
where the vector potential $A_1^{\rm ad}$ is given by 
\begin{equation}
\tilde{A}_{1z}^{\rm ad}(y) = \tilde{A}_1(0) \left(1 + \frac{{\rm tanh}(|y|/\lambda)}{k \lambda} \right) {\rm exp}( - k |y|).
\label{eq:ex-solution-tearing-A}
\end{equation}
Then, we obtain the adiabatic energy change as \citep[e.g.][]{Coppi66} 
\begin{eqnarray}
W^{\rm ad} &=& \frac{\partial}{\partial t} \int_{- \infty}^{+ \infty} \frac{\bf B_1^2}{8 \pi} dy
+\int_{- \infty}^{+ \infty}  J_{1z}^{\rm ad} \cdot E_{1z}^* dy \nonumber \\
&=& \frac{\partial}{\partial t}  \left( \frac{1}{8 \pi} \int_{- \infty}^{+ \infty} dy 
\left( |\frac{\partial}{\partial z} A_{1z}^{\rm ad}|^2+k^2|A_{1z}^{\rm ad}|^2
   -\frac{2 |A_{1z}^{\rm ad}|^2}{\lambda^2 {\rm cosh}^2(y/\lambda)} 
\right) \right)  \nonumber \\
&=& -\frac{\gamma}{2 \pi} \frac{1}{k \lambda^2} (1-k^2 \lambda^2) \tilde{A}_1(0)^2,
\label{eq:ad_energy}
\end{eqnarray}
where $\gamma=\rm{Im}(\omega)$ is the growth rate of the tearing instability.
Note that the adiabatic energy change $W^{\rm ad}$ shares with all other tearing modes, regardless of relativistic or non-relativistic reconnection \citep[e.g.][]{Coppi66,Zelenyi79}.
It can be seen that the adiabatic energy change linearly responds to the growth rate, i.e., $W^{\rm ad} \propto \gamma$.

We can likewise estimate the non-adiabatic energy change using Equation(\ref{eq:j_res}).
\begin{eqnarray}
W^{\rm res} &=& -\sum_{\pm} \int_{- l_j}^{+ lj} J_{1z}^{\rm res} \cdot E_{1z}^* dy \nonumber \\
& \simeq & -\pi \left( \frac{ \gamma \tilde{A}_1(0)}{c} \right)^2  \left( \frac{mc^2}{\Theta} \right)
\frac{\bar{N} e^2}{m} \frac{\Gamma_{\beta}^3}{kc} \lambda \left( \frac{\beta}{2} \right)^{1/2},
\label{eq:non-ad_energy}
\end{eqnarray}
where we have assumed an almost constant profile of $A_{1z}(y) \sim A_1(0)$ for $|y| < l_{\pm}$. 

We find $W^{\rm res} \propto \gamma^2$, which has the different response to the growth rate from $W^{\rm ad} \propto \gamma$.
Therefore, by likening the equations of (\ref{eq:ad_energy}) and (\ref{eq:non-ad_energy}), we were able to obtain the linear growth rate of the relativistic tearing instability as
\begin{equation}
  \gamma \tau_c \simeq \frac{2 \sqrt{2}}{\pi} k \lambda \left( 1-k^2 \lambda^2 \right)
  \left( \frac{\beta^{3/2}}{\Gamma_\beta} \right).
\label{eq:growth_rel}
\end{equation}
We have used the relationship of the Harris solution.  Note that we added the correction term of $k \lambda \sim O(1)$, because the evaluation of $W^{\rm res}$ is overestimated for $k \lambda \ll 1$ due to the assumption of the constant profile of $A_{1z}$. 

\begin{figure}
\plotone{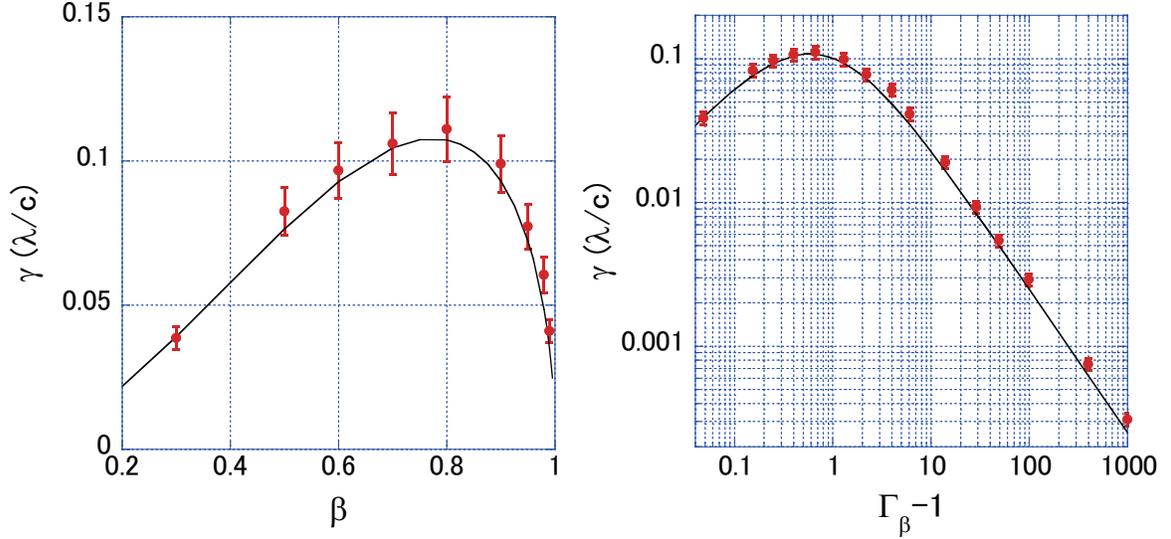}
\caption{Growth rates of relativistic tearing instability as function of the drift velocity $\beta$.  Same format as Figure 1.  The red circles are the simulation results, and the solid lines are our theoretical curve given by Eq.(\ref{eq:growth_rel}).}
\label{fig:FIG2}
\end{figure}

Based on the foregoing theoretical study, we were able to compare the generalized linear theory on relativistic tearing instability with the PIC simulation result. Fig. \ref{fig:FIG2} exhibits the comparison of the relativistic tearing instability between the generalized linear theory discussed above (solid line) and the PIC simulation (red circles). In the right-hand panel, the growth rates obtained from our simulations are extended up to $\Gamma_{\beta} = 10^3$. We found that the agreement is very good in a wide range of the drift speed $\beta c$. Moreover, we determined that the relativistic tearing instability can be suppressed in a large $\beta$ regime.

So far we assumed a relativistically hot plasma sheet with $\sigma=20$ in the PIC simulation, but the growth rate of Equation (\ref{eq:growth_rel}) can hold for a non-relativistic cold plasma sheet with a relativistic drift current, because the estimation of the non-adiabatic/resonance current in Equation (\ref{eq:j_res0}) is applicable to this case.  Note that the meandering orbit becomes larger compared to the thickness of the current sheet for $\Gamma_{\beta} \gg T/mc^2$, but the non-adiabatic/resonant current contribution is limited to inside the high density plasma sheet. Then the non-adiabatic energy $W^{\rm res}$ may not change.

\section{Discussions}

We have discussed the tearing instability that occurs in a 2D plane perpendicular to the initial electric current direction. It is, however, known that the drift-kink instability that takes place in another 2D plane including the electric current may play an important role in magnetic energy dissipation. The drift-kink instability is associated with the deformation of the current sheet into the wavy structure, thereby leading to the thickening of the current sheet \citep{Zenitani05a}. It would be important to understand the competition process between the two major instabilities for a pair plasma in a 3D system.
The growth rate of the drift-kink instability is given by
\begin{equation}
  \gamma \tau_c \simeq \frac{1}{16 \Gamma_{\beta} \beta},
\end{equation}
for a pair plasma with a relativistic temperature of $T/mc^2 \gg 1$ \citep{Zenitani07}. Furthermore, we know that the drift-kink instability is also suppressed in the regime of $\Gamma_{\beta} \gg 1$.
It would be interesting to note that the most unstable wavenumber $k$ for the drift-kink instability is given by
$k \lambda \simeq 1/(8 \Gamma_{\beta} \beta^2)$. The unstable wavenumber decreases with the increasing $\Gamma_{\beta}$ for the drift-kink mode \citep{Zenitani07}, while that of the tearing mode remains as $k \lambda \sim 0.5$ and does not depend on the drift velocity $\beta$. The characteristics of the unstable wavelength are different between the tearing mode and the drift-kink mode, but both growth rates are strongly suppressed in the relativistic current sheet with $\Gamma_{\beta} \gg 1$.

Furthermore, it may be interesting to note the simulation study about the nonlinear time evolution under the competition process between the drift-kink mode and the tearing mode in a 3D system \citep[e.g.][]{Zenitani05b,Sironi14}. \citet{Sironi14} studied the time evolution of 3D reconnection with a large $\sigma$ but with non-relativistic drift speed by using a PIC simulation code and found that the time evolution in the early time stage starts from the drift-kink instability with a periodic wavy structure along the drift current direction. It was further found that the corrugated current sheet evolves into a broader current sheet at the later stage. However, in the late nonlinear stage, the tearing mode/reconnection was found to have been initiated in association with the formation of magnetic islands. For a large $\Gamma_{\beta} \gg 1$ regime, the unstable wavelengths of the tearing mode and the drift-kink mode are different, and if anything happens, those two modes seem to be decoupled until an early nonlinear stage.

Let us make a few comments on the collisionless conductivity, which plays an important role on the magnetic field energy dissipation rate. The conductivity in the non-relativistic plasma is known to be given by
\begin{equation}
 \sigma_{\rm eff}^{\rm nonrel} \simeq \frac{N e^2}{m} \frac{1}{k v_{th}},
\end{equation}
where $v_{th}$ is the thermal velocity. By comparing the collisionless conductivity with the standard notation of conductivity $\sigma_c = (N e^2)/(m \nu_c)$, where $\nu_c$ is the collision frequency, we were able to interpret the term of $k v_{th}$ as the effective collision frequency. In particular, the effective collision in the collisionless tearing mode can be provided by the scattering of particle traveling with the thermal velocity $v_{th}$ across the tearing island of the width of $k^{-1}$. This idea is the same as Landau damping between a charge particle and an electric field. Moreover, in terms of the tearing mode, the electric field corresponds to the zero-frequency reconnection electric field around the diffusion region. If the thermal velocity $v_{\rm th}$ is small, it results in a large collisionless conductivity and the suppression of the growth of reconnection.

By comparing the foregoing non-relativistic conductivity to that of the relativistic one given in Equation (\ref{eq:sigma_rel}), we may interpret that the mass of $m$ in the non-relativistic conductivity is replaced by the relativistic inertia mass of $m (T/mc^2)$ in a relativistic hot plasma, and the effective collision frequency of $k v_{th}$ in the non-relativistic plasma becomes $k c /\Gamma_{\beta}$ by considering the relativistic retarded time effect.

Although our theoretical argument suggests that the reconnection in a pulsar wind may be suppressed during its propagation outward to the termination shock, one may think that the tearing instability/reconnection happens before the drift velocity becomes close to the speed of light.
However, we should also pay attention to the pulsar wind expanding with the relativistic speed.

It is argued that the bulk Lorentz factor $\Gamma_{\rm wind}$ is about $10^2-10^3$ at the beginning of the wind expansion, and the wind would be accelerated up to $\Gamma_{\rm wind} \sim 10^6$ \citep{Arons83,Kirk99,Lyubarsky01,Kirk03}.  It may depend on pulsar wind models and still remains to be an open question.
While the typical time scale of reconnection in the proper frame is given by $\tau_{\rm growth} > 10 (\lambda/c)$, the typical time scale in the observer frame becomes $\tau^{obs}_{\rm growth} > 10 (\varepsilon r_{lc} /c) \Gamma_{\rm wind}^2$, where the thickness of the current sheet $\lambda$ in the proper frame is $\varepsilon r_{lc}\Gamma_{\rm wind}$. 
Substituting the traveling distance of the pulsar wind before reconnection as
$r = \tau^{obs}_{\rm growth} c$ into Equation (\ref{eq:drift_vel}), we obtain $u_d/c > (5/2) (\Gamma_{\rm wind}/\kappa)$.
Then, we find the drift speed in the proper frame likely reaches up to the speed of light when $\Gamma_{\rm wind} > \kappa$.   If $\Gamma_{\rm wind} < \kappa$ at the beginning, $\Gamma_{\rm wind}$ would increase during the wind expansion due to the energy dissipation by reconnection \citep{Coroniti90,Lyubarsky01,Kirk03,Cerutti17}, and then the reconnection may cease.

We emphasized the regime of $\beta \sim 1$ during the wind expansion, but it is an open question how the current sheet of $\beta \sim 1$ is maintained.  In Equation (\ref{eq:J_drift}), we neglected the effect of the displacement current, which might play an important role on the formation of the current sheet.  
It would be important to understand the structure of the strip current sheet during the pulsar wind expansion in details in future studies.


\end{document}